\documentclass{article}

\PassOptionsToPackage{numbers, compress}{natbib}
\usepackage[nonatbib, final]{neurips_2024}

\usepackage[utf8]{inputenc}
\usepackage[T1]{fontenc}
\usepackage{xurl}
\usepackage[colorlinks, allcolors=black]{hyperref}
\usepackage{booktabs}
\usepackage{amsfonts}
\usepackage{nicefrac}
\usepackage{microtype}
\usepackage{xcolor}
\usepackage{graphicx}
\usepackage{tabularx}
\usepackage{enumitem}
\usepackage{multirow}
\usepackage{array}
\usepackage{apacite}
\usepackage{cleveref}
\let\citep\shortcite
\let\citet\shortciteA
\let\citealp\shortciteNP

\title{Risk thresholds for frontier AI}

\author{
Leonie Koessler\thanks{Corresponding author: \url{leonie.koessler@governance.ai}.} \quad
Jonas Schuett \quad
Markus Anderljung \\ \\
Centre for the Governance of AI 
}

\begin{document}
\maketitle
\setcounter{footnote}{0}

\begin{abstract}
Frontier artificial intelligence (AI) systems could pose increasing risks to public safety and security. But what level of risk is acceptable? One increasingly popular approach is to define \emph{capability thresholds}, which describe AI capabilities beyond which an AI system is deemed to pose too much risk. A more direct approach is to define \emph{risk thresholds} that simply state how much risk would be too much. For instance, they might state that the likelihood of cybercriminals using an AI system to cause X amount of economic damage must not increase by more than Y percentage points. The main upside of risk thresholds is that they are more principled than capability thresholds, but the main downside is that they are more difficult to evaluate reliably. For this reason, we currently recommend that companies (1) define risk thresholds to provide a principled foundation for their decision-making, (2) use these risk thresholds to help set capability thresholds, and then (3) primarily rely on capability thresholds to make their decisions. Regulators should also explore the area because, ultimately, they are the most legitimate actors to define risk thresholds. If AI risk estimates become more reliable, risk thresholds should arguably play an increasingly direct role in decision-making.
\vspace{1.6em}

\begin{figure}[h!]
    \makebox[\textwidth][c]{\includegraphics[width=1.115\linewidth]{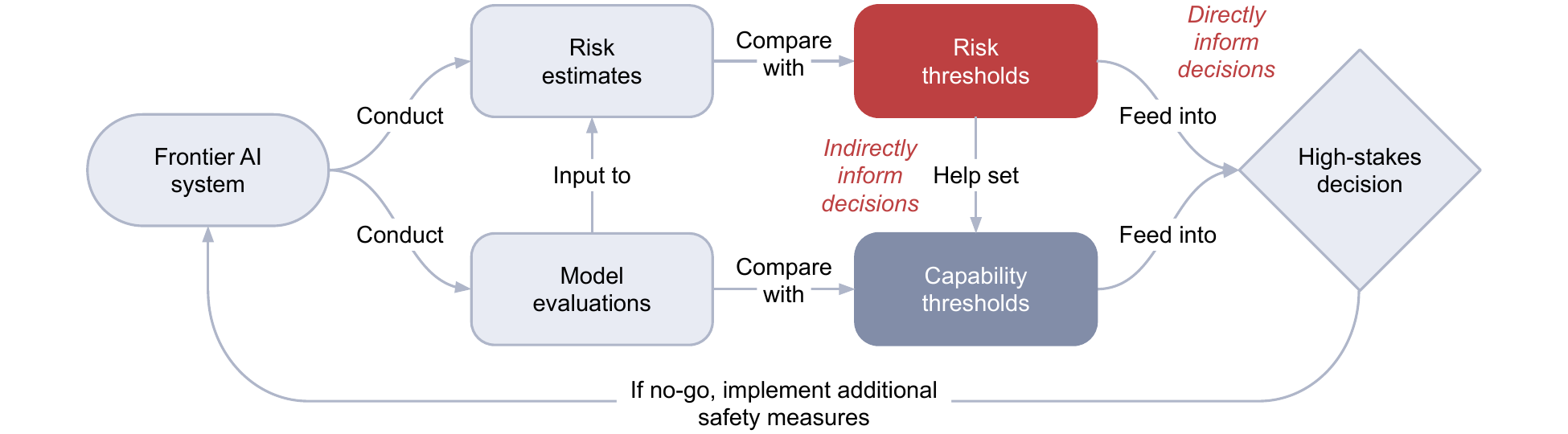}}
    \caption{Risk thresholds can directly and indirectly inform high-stakes AI development and deployment decisions.}
    \label{fig:1}
    \end{figure}
\end{abstract}

\newpage
\tableofcontents
\newpage

\section*{Executive summary}

Frontier artificial intelligence (AI) systems could pose increasing risks to public safety and security (e.g. through cyberattacks on critical infrastructure, the acquisition of biological weapons, or loss of control over AI systems). These risks could largely stem from a small number of high-stakes development and deployment decisions made by frontier AI companies (e.g. whether to start a large training run or whether to release a model). When making such decisions, companies do not seem to use risk thresholds, i.e. limits for what likelihood and severity of harm they are willing to accept. Instead, where companies have defined thresholds for what AI systems are too risky to release, those thresholds have been defined in terms of model capabilities. This paper draws on other industries to discuss how to use risk thresholds for making high-stakes AI development and deployment decisions.

\textbf{Risk thresholds serve a different function than capability thresholds and compute thresholds (\Cref{risk-thresholds-and-related-concepts}).}

\begin{itemize}[leftmargin=2em]
    \item \textbf{Compute thresholds.} Compute thresholds are defined in terms of computational resources used to train a model (``training compute''). Training compute is a very imperfect proxy for risk, but can easily be measured and forecasted early on in the development process. Compute thresholds should thus be used as an initial filter to identify models that warrant further scrutiny, oversight, and precautionary safety measures.

    \item \textbf{Capability thresholds.} Model capabilities are a better proxy for risk than training compute and are easier to evaluate than risk. Capability thresholds may therefore serve as a key trigger for whether additional safety measures should be implemented before a high-stakes activity may go ahead.

    \item \textbf{Risk thresholds.} Risk estimates try to measure the level of risk directly, but they are still highly unreliable. In theory, risk thresholds are the ideal determinator for when additional safety measures are necessary. But in practice, risk thresholds cannot yet be relied upon for decision-making. More on the role they should play below.
\end{itemize}

\textbf{In principle, there are two ways in which risk thresholds can be used: they can directly feed into high-stakes AI development and deployment decisions and they can indirectly feed into such decisions by helping set capability thresholds (\Cref{how-to-use-ai-risk-thresholds}).} These two ways are illustrated in \Cref{fig:1}.

\begin{itemize}[leftmargin=2em]
    \item \textbf{Directly feeding into decisions.} Using risk thresholds to directly feed into high-stakes decisions is the most common use case for risk thresholds in other industries. Before making a high-stakes decision, many companies compare risk estimates to predefined risk thresholds. If the estimated level of risk is above the risk thresholds, companies implement additional safety measures and repeat the process. This process is similar to how some frontier AI companies evaluate model capabilities and compare them to predefined capability thresholds, but with a focus on risk rather than model capabilities. In this way, both risk thresholds and capability thresholds can directly feed into high-stakes decisions.

    \item \textbf{Indirectly feeding into decisions.} Using risk thresholds to indirectly feed into decisions is less common in other industries. One exception are U.S. nuclear regulators who use risk thresholds to determine adequate safety measures. In the context of frontier AI, capability thresholds and corresponding safety measures could be designed such that they would be estimated to keep risk below some risk thresholds. To that end, risk thresholds need to be defined. Next, risk models can be developed, i.e. mappings of pathways from risk factors to harm. These risk models can help identify the model capabilities at which risk would exceed the risk thresholds, and the safety measures that would keep risk below the risk thresholds. The identified model capabilities then serve as the capability thresholds that trigger the identified safety measures.
\end{itemize}

\textbf{We argue that risk thresholds are a promising tool for frontier AI regulation (\Cref{the-case-for-ai-risk-thresholds}).}

\begin{itemize}[leftmargin=2em]
    \item \textbf{Arguments for using risk thresholds.} Risk thresholds may help align business conduct with societal concern; enable consistent allocation of safety resources; ensure risk estimation results are actually acted upon; prevent motivated reasoning regarding what level of risk is acceptable; and avoid locking in premature safety measures.

    \item \textbf{Arguments against using risk thresholds.} Risk thresholds rely on risk estimates but estimating risks from AI is extremely hard; AI is a dual-use, general-purpose technology; risk thresholds may create an incentive to produce artificially low risk estimates; and defining risk thresholds for AI involves handling thorny normative trade-offs.

    \item \textbf{How risk thresholds should be used.} Overall, we suggest that risk thresholds should be used to indirectly feed into decisions by helping set capability thresholds. Yet risk thresholds should only inform, but not determine, where to set capability thresholds: risk thresholds should not be the sole basis of a strict decision-rule. Other considerations should also be taken into account when setting capability thresholds. Further, we suggest that risk thresholds may be used to directly feed into decisions. However, again, risk thresholds should only inform decisions (e.g. as one of a number of considerations), and not determine decisions (e.g. as the sole basis for a strict decision-rule). If and when our ability to produce risk estimates improves, we can rely more on risk thresholds.
\end{itemize}

\textbf{Finally, we propose a framework for how to define risk thresholds for frontier AI (\Cref{how-to-define-ai-risk-thresholds}).} Before regulators or companies can answer the question of what level of risk is acceptable, they need to decide which type of risk the threshold should refer to, that is, which risk scenarios are in scope. Next, when determining the acceptable level of risk, they need to handle three related normative trade-offs: (1) how to weigh potential harms and benefits, (2) to what extent should mitigation costs be taken into account, and (3) how to deal with uncertainty regarding all of the aforementioned.

We encourage frontier AI companies to start experimenting with risk thresholds today. Regulators should also explore the area because, ultimately, they are the most legitimate actors to define risk thresholds. To this end, we need a discussion about what level of risk we, as a society, are willing to accept.
\newpage

\section{Introduction}\label{introduction}

Frontier artificial intelligence (AI) systems\footnote{We define ``frontier AI systems'' as ``highly capable general-purpose AI models or systems that can perform a wide variety of tasks and match or exceed the capabilities present in the most advanced models'' \citep{Dsit2024-bk}. For example, this currently includes systems like GPT-4, Claude 3, and Gemini Ultra. Note that, in contrast to an earlier, otherwise identical definition \citep{Dsit2023-dv}, this definition has replaced ``today's most advanced models'' with ``the most advanced models'', which implies that the frontier changes as models become more capable. We also note that the term ``frontier AI'' has been accused of promoting a specific worldview \citep{Helfrich2024-av}.} pose increasing risks\footnote{We define ``risk'' as the combination of likelihood and severity of harm \citep{Iso2014-va}. A recent trend in risk management uses a definition of risk that includes both negative impacts, i.e. harm, and positive impacts, i.e. benefits \citep{Coso2017-ww,Iso2018-xt,Nist2023-nr}. However, in the context of risk thresholds, the understanding of risk typically only includes harm, whereas benefits come into play as the key consideration when choosing what level of risk is acceptable (see \Cref{level-of-risk}).} to public safety and security\footnote{Note that for the purposes of this paper, we focus on risks to individuals, groups, and society as a whole, i.e. societal risks (e.g. fatalities, economic damage, and societal disruption). This means we ignore risks to the company itself, i.e. business risks (e.g. financial risks, legal risks, and reputational risks). We also focus on risks to public safety and security, but the tools we discuss can likely be applied to many other types of societal harm, too.} \citep{Bengio2024-ca,Hendrycks2023-oz,Anderljung2023-pa}. For example, frontier AI systems may already increase cybercriminal productivity \citep{Fang2024-mc,Hazell2023-bx,Lohn2022-kn,Mirsky2021-sa}, while future systems might increase the risk that terrorists will succeed in acquiring biological weapons \citep{Boiko2023-nm,Mouton2023-tt,Sandbrink2023-xn,Soice2023-ty,Urbina2022-qa}. A more speculative concern is that, at some point, frontier AI systems might evade human control and cause large-scale harm on their own \shortcite{Chan2023-aj,Cohen2024-jl,Hendrycks2023-oz,Ngo2024-fx}.

Theses risks could largely stem from a small number of high-stakes
development and deployment decisions made by frontier AI companies, such
as whether to start a final large training run or whether to deploy a model, also 
referred to as ``go/no-go decisions'' \citep{Nist2023-nr}. When making these 
decisions, companies necessarily accept some level of risk.\footnote{We define
``level of risk'' as the combined measure of the likelihood and
severity of harm.} For example, a company deploying a system could be
accepting a 0.01\% increase in the risk that a malicious actor will
succeed in acquiring a biological weapon based on instructions from that
system. Many frontier AI companies seem to consider potential harms and
benefits to society in their decision-making (e.g. 
\citealp{Anthropic2023-bl,Google2018-dj,Google_DeepMind2024-qj,Meta2023-gj,Microsoft2024-qn,OpenAI2023-tt}).
However, companies do not appear to have clear limits for what
likelihood and severity of harm they are willing to accept, so-called
``risk thresholds''.\footnote{\citet{Google2018-dj}
commits that it ``will not design or deploy AI (...) applications
{[}that{]} cause or are likely to cause overall harm. Where there is a
material risk of harm, we will proceed only where we believe that the
benefits substantially outweigh the risks, and will incorporate
appropriate safety constraints.'' This can be understood as a risk
threshold, albeit a very vague one. Much depends on how Google
operationalizes this risk threshold.}

At the 2024 AI Summit in South Korea, governments and companies both
emphasized the importance of setting thresholds above which risk would be 
unacceptable. The Seoul Ministerial
Statement includes the intention to ``identify thresholds at which the
level of risk posed by the design, development, deployment and use of
frontier AI models or systems would be severe absent appropriate
mitigations'' \citep{Dsit2024-an}. The Seoul Frontier AI
Safety Commitments had 16 companies commit to ``set out thresholds at
which severe risks posed by a model or system, unless adequately
mitigated, would be deemed intolerable'', while noting that ``thresholds
can be defined using model capabilities, estimates of risk {[}i.e.
``risk thresholds''{]}, implemented safeguards, deployment contexts
and/or other relevant risk factors'' \citep{Dsit2024-bk}.
While the past year has seen frontier AI companies increasingly define 
thresholds in terms of model capabilities, it is unclear whether these 
thresholds keep risk to an acceptable level. This paper draws on other 
industries to discuss how regulators and companies should use risk thresholds for making high-stakes AI development and deployment decisions.

There is an extensive body of literature on risk thresholds in other
industries. Technical standards provide high-level guidance on how to
use risk thresholds for business risks (e.g.
\citealp{Coso2017-ww,Iso2018-xt,Iso2019-kp}). The
scholarly literature provides more in-depth guidance for business and
societal risks (e.g.
\citealp{Aven2012-ue,Aven2015-ug,Popov2021-ow,Rausand2020-ao})
and discusses various issues with risk thresholds, including substantial
uncertainties in risk estimates (e.g.
\citealp{Fischhoff1984-ba,Klinke2002-bm,Starr1969-sw}).
Regulators in many safety-critical industries mandate or recommend
specific risk thresholds, such as in the nuclear
\citep{Anvs2020-rn,Iaea2005-se,Nrc1983-oo}, maritime
\citep{Imo2018-bo}, aviation
\citep{Eurocontrol2001-cf,Faa1988-vk,Icao2018-hl}, and
space industries
\citep{Esa2023-tk,Faa2016-tb}.
In addition to a large corpus of industry-specific literature, many
reports survey the use of risk thresholds across industries and
jurisdictions (e.g.
\citealp{Ccps2009-rs,Ehrhart2020-dk,Flamberg2016-fe,Linkov2011-cc,Marhavilas2021-bg}).

By contrast, in the context of frontier AI development and deployment,
regulators and scholars are only starting to discuss risk thresholds.
The NIST AI Risk Management Framework recommends that companies define
``risk tolerances'' \citep{Nist2023-nr}, but does not
provide much guidance for how to define or use them. DSIT's policy paper
Emerging Processes for Frontier AI Safety recommends that companies use
risk thresholds in responsible capability scaling
\citep{Dsit2023-xm}, but it only provides high-level
guidance. Further, the forthcoming EU AI Act mandates that risk
management measures for general-purpose AI models with systemic risk
``shall be proportionate to the risks {[}and{]} take into consideration
their severity and probability'' (Article~56(2)(d)). This could be
ensured by using risk thresholds.\footnote{Similarly, risk management
measures for high-risk AI systems ``shall be such that the relevant
residual risk associated with each hazard, as well as the overall
residual risk of the high-risk AI systems is judged to be acceptable''
(Article~9(5)). For related discussions, see
\citep{Fraser2023-oa,Laux2024-hi}, and
\citep{Schuett2023-be}. Moreover, the EU AI Act puts AI systems into risk 
categories, the boundaries of which have been referred to as risk thresholds 
\citep{Novelli2024-ho}, although they are not defined in terms of likelihood and 
severity of harm, and therefore do not qualify as risk thresholds according to our 
definition.} There is only tangential scholarly treatment of AI risk thresholds
\citep{Clymer2024-sr}. Taken together, there is a clear
need for more concrete guidance on how to use risk thresholds in the
context of frontier AI. This paper aims to help fill this gap.

The paper proceeds as follows. First, we introduce the concept of risk
thresholds as a specific type of risk acceptance criteria and
differentiate it from the related concepts emerging in the frontier AI
context: capability thresholds and compute thresholds (\Cref{risk-thresholds-and-related-concepts}).
We then outline how risk thresholds can be used to directly and
indirectly feed into high-stakes AI development and deployment decisions
(\Cref{how-to-use-ai-risk-thresholds}).
Next, we argue that risk thresholds should only be used to inform, but not
determine, high-stakes decisions, unless risk estimates become more
reliable
(\Cref{the-case-for-ai-risk-thresholds}).
We also highlight key considerations and provide initial guidance for
defining AI risk thresholds
(\Cref{how-to-define-ai-risk-thresholds}).
We conclude with a summary of our main contributions and suggestions for
further research (\Cref{conclusion}).

\section{Risk thresholds and related 
concepts}\label{risk-thresholds-and-related-concepts}

In frontier AI regulation, different thresholds are currently emerging:
risk thresholds, capability thresholds, and compute thresholds. These
thresholds are predefined values above which additional safety measures
are deemed necessary. The thresholds differ regarding the \emph{metric}
in terms of which they are defined (risk, model capabilities, and
training compute), and the \emph{function} they serve in frontier AI
regulation (we discuss this for each threshold below). For example,
Anthropic's Responsible Scaling Policy maps specific model capabilities
to specific safety measures \citep{Anthropic2023-bl},
whereas the EU AI Act classifies general-purpose AI models trained on
more than $10^{25}$ floating-point operations as posing
systemic risk (Article~51(2)) and imposes more stringent requirements on
their providers (Article~55(1)).

In this paper, we are most interested in thresholds that can be used in
high-stakes decision-making to determine whether the risk from the
development and deployment of a frontier AI system is acceptable. This
includes both risk thresholds and capability thresholds, though we will
focus on risk thresholds based on risk estimates (see
\citealp{Dsit2024-bk}).

In the remainder of this section, we first conceptualize risk thresholds
as risk acceptance criteria and outline how they are used in other
industries to directly and indirectly feed into high-stakes decisions
(\Cref{risk-thresholds}). We then
argue that capability thresholds can also be considered risk acceptance
criteria that may serve as a key trigger for when to implement
additional safety measures in the frontier AI context
(\Cref{capability-thresholds}).
Finally, we assert that compute thresholds should not be considered risk
acceptance criteria but only serve as an initial filter to identify
models of potential concern
(\Cref{compute-thresholds}).

\subsection{Risk thresholds}\label{risk-thresholds}

Risk thresholds are limits to what level of estimated risk is acceptable
\citep{Aven2015-ug}. Thus, they are also referred to as
``risk limits'' \citep{Iso2019-kp}, ``tolerability
limits'' \citep{Aven2015-ug}, or ``risk tolerances''
\citep{Nist2023-nr}. In the context of business risks,
risk thresholds are also sometimes referred to as companies' ``risk
appetite'' \citep{Coso2017-ww,Iso2019-kp}. Risk
thresholds vary across different industries and jurisdictions
\citep{Ehrhart2020-dk,Flamberg2016-fe,Linkov2011-cc}. For
example, in the U.S. aviation industry, the probability of ``failure
conditions which would prevent continued safe flight and landing''
should not exceed $1\times 10^{-9}$ (one in a billion) per
flight-hour \citep{Faa1988-vk}. As another example, in
the UK nuclear industry, the risk of death of a member of the public is
``unacceptable'' if it is above $1\times 10^{-4}$ per
plant-year and ``broadly acceptable'' if it is below
1~×~10\textsuperscript{-6} per plant-year
\citep{Onr2020-yu}.

Risk thresholds can be understood as a particular type of ``risk
acceptance criteria'', i.e. criteria that establish the conditions under
which risk is acceptable to an organization (e.g. a regulator or a
company). Therefore, risk acceptance criteria are also referred to as
``risk evaluation criteria'', ``decision criteria for risk management
decision making'', or simply ``risk criteria''
\citep{Aven2012-ue,Aven2015-ug,Aven2016-fj,Iso2018-xt,Iso2019-kp,Morgan1990-ts}.\footnote{Note
that concepts and terminology vary among sources or are simply
unclear. Some authors seem to equate risk thresholds with risk
acceptance criteria (e.g. \citealp{Linkov2011-cc}),
whereas other authors seem to understand risk thresholds as
quantitative risk acceptance criteria (e.g.
\citep{Flamberg2016-fe}). For most authors, it simply
remains unclear how they conceptualize the relationship between risk
thresholds and risk acceptance criteria.} Risk acceptance criteria
beyond risk thresholds can take many forms. For example, risk may be
acceptable if it is ``as low as reasonably practicable'' (``ALARP''), if
the ``best available technology'' (``BAT'') is used, or if the affected
individuals have given consent
\citep{Klinke2002-bm,Morgan1990-ts,Vanem2012-vs}.
Compared to other types of risk acceptance criteria, risk thresholds are
more often quantitative, although they can also be qualitative (e.g.
``only proceed if risk is deemed low''). However, in the regulatory
context, qualitative risk thresholds appear to be very uncommon.

We highlight that choosing a type of risk acceptance
criteria may reflect a particular ethical viewpoint. Although this
viewpoint can significantly affect which risks are deemed acceptable, it
is rarely made explicit. Common ethical principles that may underlie
different types of risk acceptance criteria include principles of
utility, fairness, and human rights
\citep{Morgan1990-ts,Vanem2012-vs}. Risk thresholds may
draw most strongly on the principle of utility, because they focus on
potential harms and benefits, outcomes rather than processes, and
general welfare rather than individual liberties.\footnote{On the
relationship between utility and rights, see e.g.
\citep{Hart2017-ru}.} However, other principles can be
taken into account via the design of the risk thresholds (see
\citealp{Morgan1990-ts,Vanem2012-vs}). For example, U.S.
oil and gas facilities have to observe stricter risk thresholds
regarding particularly vulnerable groups in places such as schools,
hospitals, and prisons \citep{Nfpa2023-yi}. Furthermore,
participatory elements can be included when setting risk thresholds, for
instance, through public consultations (e.g.
\citealp{Nrc1983-oo}). Finally, we do not argue that risk
thresholds should be the only risk acceptance criteria in the frontier
AI context.

\begin{figure}
    \centering
    \includegraphics[width=\linewidth]{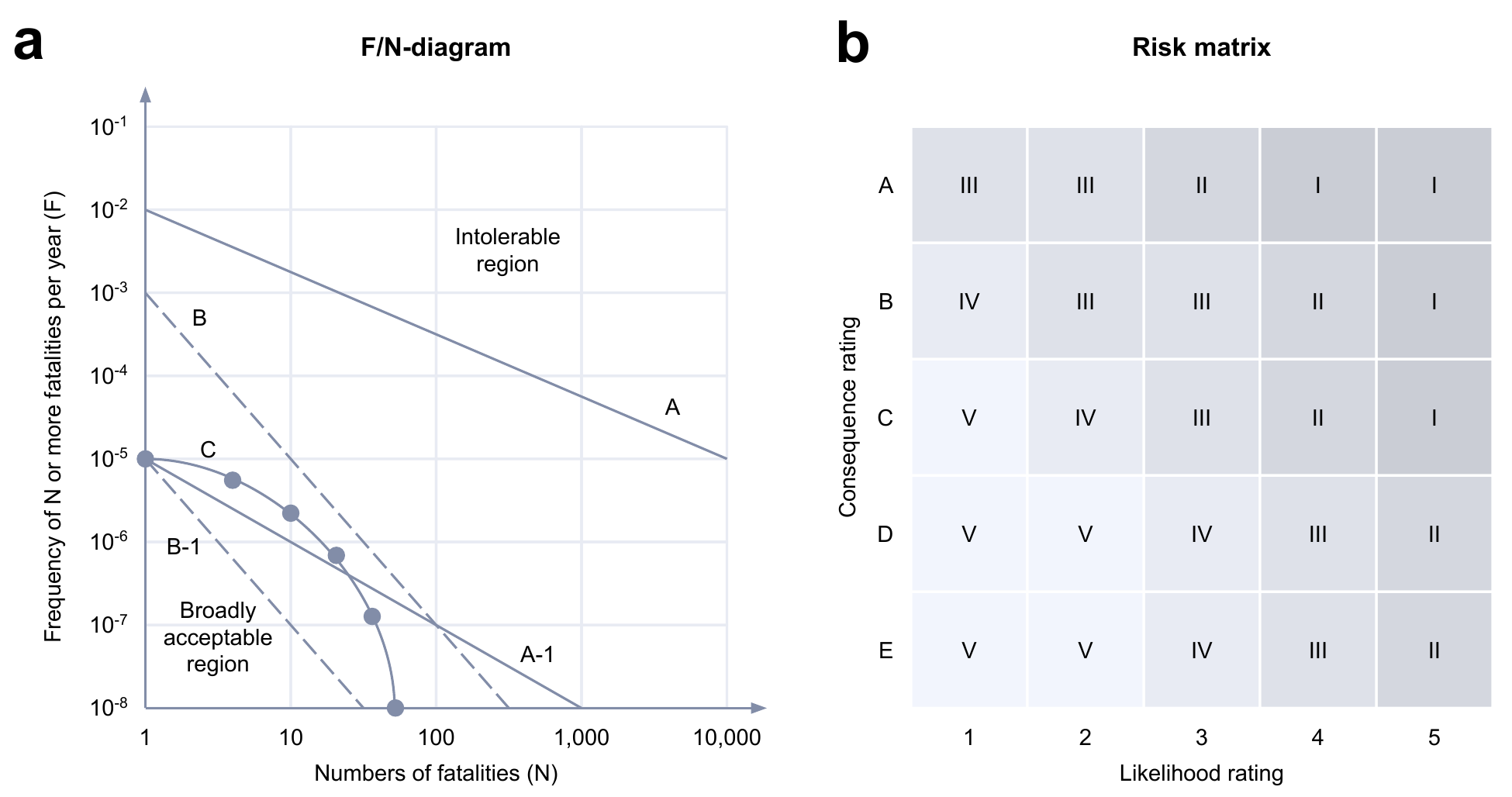}
    \caption{F/N-diagram (quantitative) and risk matrix (semi-quantitative\,/\,qualitative) \protect\citep{Iso2019-kp}}
    \label{fig:2}
\end{figure}

Risk thresholds are defined in terms of likelihood and severity of harm.
Likelihood scales refer to the probability of events, which can be
estimated using historical data, models, or expert judgment, among other
things. Severity scales refer to the magnitude or degree of some type of
harm, such as fatalities, injuries, or economic damage. They can also be
defined in terms of potentially harmful events, such as a successful
cyberattack, the acquisition of a biological weapon, or the creation of
a deepfake.\footnote{In this paper, for simplicity, we focus on events
that are intrinsic harms. On the one hand, the likelihood of
potentially harmful events will usually be easier to estimate than the
likelihood of intrinsic harms. On the other hand, the question of at
what level to set the threshold is even more complicated for
potentially harmful events than it already is for intrinsic harms
(\Cref{level-of-risk}).} Both
likelihood and severity scales can be quantitative (i.e. numeric values,
e.g. probabilities or numbers of fatalities), semi-quantitative (i.e.
ranges of numeric values, e.g. $1-5\%$ or $10,000$-$100,000$ fatalities), or
qualitative (i.e. categories based on non-numeric values, e.g.
``likely'' or ``severe'') \citep{Iso2019-kp}. Risk
thresholds consist of a single pair of likelihood and severity values
(e.g. an expected value) or several pairs of likelihood and severity
values (e.g. a probability distribution). The latter seems to be much
more common in the regulatory context, at least for fatalities (e.g.
\citealp{Eurocontrol2001-cf,Hse2001-lo,Nrc1983-oo}).
Quantitative risk thresholds can be visualized in graphs (e.g. F/N
diagrams with fatalities N on the $x$-axis and frequencies F on the
$y$-axis), whereas semi-quantitative and qualitative risk thresholds can
be visualized in risk matrices (\Cref{fig:2}).

Risk thresholds can feed into high-stakes decisions in two ways:
\emph{directly} and \emph{indirectly}. First, when companies make
high-stakes decisions, risk thresholds can be used to help decide
whether an activity may go ahead \citep{Iso2018-xt}. In this
way, risk thresholds \emph{directly} feed into high-stakes decisions. This
is the most common way in which other industries use risk thresholds.
Second, instead of using risk thresholds on a case-by-case basis, risk
thresholds can also be used to help specify which safety measures need
to be implemented under which circumstances. In this way, risk
thresholds \emph{indirectly} feed into high-stakes decisions. In the U.S.
nuclear industry, ``safety goals (...) are to be used (...) in making
regulatory judgments on the need of proposing and backfitting new
generic requirements on nuclear power plant licensees''
\citep{Nrc2021-fe}. Similarly, Anthropic evaluates for
``capability improvements (...) {[}that{]} would significantly increase
the risk (...) past an unacceptable threshold'' to decide when
additional safety measures are necessary
\citep{Anthropic2023-bl}. We elaborate on how to use risk
thresholds in the frontier AI context in
\Cref{how-to-use-ai-risk-thresholds}.

\subsection{Capability thresholds}\label{capability-thresholds}

For risks to public safety and security, model capabilities can be
considered a key risk factor and even an imperfect proxy for risk.
Fundamentally, risks from frontier AI stem from the capabilities a model
possesses, because many of these capabilities are dual-use: they can be
used for good or for evil
\citep{Anderljung2023-xo,Bommasani2021-rj,Shevlane2020-kc}.
For example, a model that can be used by scientists to help develop new
pharmaceuticals might also be used by terrorists to help develop new
toxins \citep{Urbina2022-qa}. Thus, model capabilities
can be considered a key risk factor. They can even be considered a proxy
for risk, a claim that has been made explicitly by some (e.g.
\citealp{Sastry2024-sf}) and to some extent implicitly
relied upon by others
\citep{Anderljung2023-pa,Shevlane2023-ov,OpenAI2023-tt}.
But factors other than model capabilities are crucial for risk too, such
as the number, capacity, and willingness of malicious actors to use the
model or the level of societal preparedness
\citep{Anderljung2023-pa,Bernardi2024-kn,Kapoor2024-ss}.
Nevertheless, a model's capabilities are easier to evaluate than its risk
(\Cref{arguments-against-using-risk-thresholds}),
making them a useful metric for frontier AI regulation.

Already, frontier AI companies increasingly rely on capability
thresholds to make high-stakes development and deployment decisions. 
Capability thresholds are predefined model capabilities at
which additional safety measures are deemed necessary. Three frontier AI
companies have published policies that define capability thresholds and
when additional safety measures should be implemented before these
capability thresholds are crossed. This includes
Anthropic's Responsible Scaling Policy
\citep{Anthropic2023-bl}, OpenAI's
Preparedness Framework \citep{OpenAI2023-tt}, and Google
DeepMind's Frontier Safety Framework
\citep{Google_DeepMind2024-qj}. These policies focus on
chemical, biological, radiological, and nuclear (CBRN); cyber;
persuasion; autonomy; and some other capabilities, measured by so-called
``model evaluations''
\citep{Shevlane2023-ov,Phuong2024-qu}. Regulators have
yet to make use of capability thresholds, but some of them already seem
to be thinking along these lines \citep{Dsit2023-xm}.

While concepts in the frontier AI context are still evolving, capability
thresholds can be considered their own type of risk acceptance criteria.
Capability thresholds essentially define conditions under which a risky
activity may go ahead, namely if a model's capabilities are below the
threshold or if they are above the threshold but adequate safety
measures have been implemented. In this way, capability thresholds can
be considered a type of risk acceptance criteria that is distinct from
risk thresholds. However, concepts in the frontier AI context are still
evolving. In particular, not all model evaluations only measure inherent
model capabilities; they may also include assessments of how users or
even society as a whole interact with models
\citep{Dsit2024-qc,Patwardhan2024-ep,Solaiman2023-yt}.
Moreover, risk thresholds can be used to help with setting capability
thresholds, a process that blurs the lines between risk thresholds and
capability thresholds.\footnote{Indeed, OpenAI refers to its capability
thresholds as ``risk thresholds'' \citep{OpenAI2023-tt}
-- presumably because it aims for its capability thresholds to keep
risk at an acceptable level. However, OpenAI does not define its
thresholds in terms of likelihood and severity of harm, but model
capabilities. Therefore, according to our definitions, these
thresholds are capability thresholds, and not risk thresholds.} We get
back to this in
\Cref{using-risk-thresholds-to-indirectly-inform-decisions}.

\subsection{Compute thresholds}\label{compute-thresholds}

\begin{figure}[t!]
    \centering
    \makebox[\textwidth][c]{\includegraphics[width=1.15\linewidth]{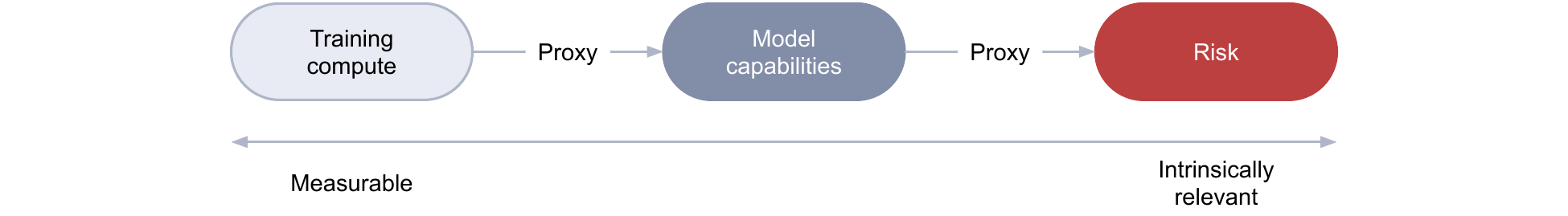}}
    \caption{Different metrics and the relationships between them}
    \label{fig:3}
\end{figure}

Under the current deep learning paradigm, the amount of computational
resources used to train a model (``training compute'') can be considered a very
imperfect proxy for a model's risk. Empirically, recent advances in
model capabilities to a large extent stem from increasing amounts of
computational resources being used to train the model, a phenomenon also
referred to as ``scaling laws''
\citep{Sutton2019-fi,Kaplan2020-uz,Hernandez2021-kr,Hoffmann2022-jr}.
While how long scaling laws will hold is somewhat contentious
\citep{Lohn2022-uu,Villalobos2022-wl}, training compute can be, at
least currently, considered a proxy for a model's capabilities and
thereby also the model's risk
\citep{Anderljung2023-pa,Sastry2024-sf}. However, training compute
is even further removed from risk than model capabilities -- they
already are an imperfect proxy for risk -- meaning that training compute is 
only a very imperfect proxy for risk. Still, training compute is relatively 
easy to measure, making it a useful metric to build on in frontier AI 
regulation
\citep{Anderljung2023-pa,Heim_at_al_undated-zb,Pistillo_et_al_undated-dl,Sastry2024-sf}.
We show the relationship between the three metrics in \Cref{fig:3}.

Indeed, regulators in the U.S. and the EU already make use of
\emph{compute thresholds} to identify models that might be of concern
and require increased scrutiny, oversight, and precautionary security
measures. The U.S. Executive Order on the Safe, Secure, and Trustworthy
Development and Use of Artificial Intelligence imposes requirements on
companies developing and deploying models above a training compute
threshold of 10\textsuperscript{26} operations to notify the government
before development; report on ownership and possession of model weights
and measures taken to secure them; and report on the results of
red-teaming tests and measures taken based on them (Section~4.2(i)).
Setting a lower threshold while imposing more extensive requirements,
the EU AI Act uses a compute threshold of 10\textsuperscript{25}
floating-point operations to identify ``general-purpose AI models'' that
may pose ``systemic risk'' (Article~51(2)), and requires providers of
such models to conduct model evaluations; assess and mitigate systemic
risks; track, document, and report serious incidents; and ensure an
adequate level of cybersecurity for the model and its physical
infrastructure (Article~55(1)).

Compute thresholds should not be used as risk acceptance criteria to
directly feed into high-stakes decisions. Because compute thresholds are
such an imperfect proxy for risk, they should not be used to define
conditions under which a risky activity may go ahead. Instead, compute
thresholds may serve as an initial filter for further scrutiny,
oversight, and some precautionary safety measures 
\citep{Heim_at_al_undated-zb,Pistillo_et_al_undated-dl}. Capability thresholds and
risk thresholds can then be used to help decide whether high-stakes
decisions may go ahead and under which circumstances additional safety
measures are warranted. The U.S. Executive Order on AI and the EU AI Act 
laudably use compute thresholds mainly in this way. Overall, risk 
thresholds, capability thresholds, and compute thresholds are not 
substitutes for each other; each has a distinct function in frontier AI 
regulation
(\Cref{tab:1})2

\renewcommand{\arraystretch}{1.2}
\begin{table}[t!]
    \centering
    \begin{tabularx}{\linewidth}{*{3}{>{\raggedright\arraybackslash}X}}
        \toprule
        \textbf{Compute thresholds} & \textbf{Capability thresholds} & \textbf{Risk thresholds}\\
        \midrule
        Initial filter for further scrutiny, oversight, and precautionary safety measures (e.g. security)
        &
        Key trigger for when additional safety measures are necessary, including fast responses (e.g. pausing)
        &
        Ideal, though immature, determinator for when additional safety measures are necessary; risk thresholds can \textit{directly} feed into high-stakes decisions and \textit{indirectly} feed into high-stakes decisions by helping set capability thresholds
        \\
        \bottomrule
        \\[-5pt]
    \end{tabularx}
    \caption{Different thresholds and their functions in frontier AI regulation}
    \label{tab:1}
\end{table}

\section{How to use AI risk
thresholds}\label{how-to-use-ai-risk-thresholds}

In this section, we discuss two ways in which risk thresholds can be
used: to directly feed into high-stakes AI development and deployment
decisions
(\Cref{using-risk-thresholds-to-directly-inform-decisions})
and to indirectly feed into decisions by helping set capability thresholds
(\Cref{using-risk-thresholds-to-indirectly-inform-decisions}).
These two use cases are illustrated in
\Cref{fig:1}.

\subsection{Using risk thresholds to directly feed into
decisions}\label{using-risk-thresholds-to-directly-inform-decisions}

Using risk thresholds to directly feed into high-stakes decisions is the
most common use case for risk thresholds in other industries
(\Cref{risk-thresholds}). In the
standard risk management process, organizations estimate the level of
risk (``risk analysis'') and compare the results to predefined risk
thresholds (``risk evaluation''). If the estimated level of risk is
above the risk threshold, companies need to implement additional safety
measures and repeat the process \citep{Iso2018-xt}. This
process is similar to how some frontier AI companies evaluate model
capabilities and compare them to predefined capability thresholds
(\Cref{capability-thresholds}), but
with a focus on risk rather than model capabilities. In this 
way, both risk thresholds and capability thresholds can directly feed into
high-stakes decisions
(\Cref{fig:1}).

If the estimated level of risk exceeds a risk threshold, the company
needs to implement additional safety measures. In the simplest version
of risk thresholds, the company can freely choose among safety measures
as long as it brings the level of risk below the risk threshold before
proceeding. But risk thresholds can also require companies to take more
specific safety measures. For example, companies may take measures to
reduce risk or uncertainty about risk as much as possible, or they may
notify some internal or external stakeholder. In the case of frontier
AI, risk thresholds could also require companies to notify the board of
directors or the competent regulator (who may be allowed to veto the
decision or be required to give permission to go ahead), or to conduct
an extra suite of in-depth model evaluations (which may have to include
external parties). Using safety measures other than a clear ``no-go''
also provides a way for risk thresholds to inform, but not determine,
high-stakes decisions. We will get back to this in
\Cref{overall-suggestions-for-using-ai-risk-thresholds}.

It is possible to set a single risk threshold or multiple risk
thresholds that trigger different safety measures. The simplest approach
is to set a single risk threshold that distinguishes two risk tiers. If
this risk threshold is crossed, the activity in question may not go
ahead unless either the risk has been reduced or specific safety
measures have been taken. But it is also possible to set multiple risk
thresholds at different levels of risk to distinguish between more than
two risk tiers. For example, two risk thresholds could distinguish
between risk being unacceptable under any circumstances, risk being
acceptable if specific safety measures have been taken (e.g. risk has
been reduced as much as possible, specific information has been
gathered, or specific internal or external actors have been notified),
and risk being acceptable without further safety measures. Stacking
multiple risk thresholds in this way allows predefining more
fine-grained decision rules for different levels of risk compared to a
single risk threshold.

\begin{figure}[t!]
    \centering
    \makebox[\textwidth][c]{\includegraphics[width=1.15\linewidth]{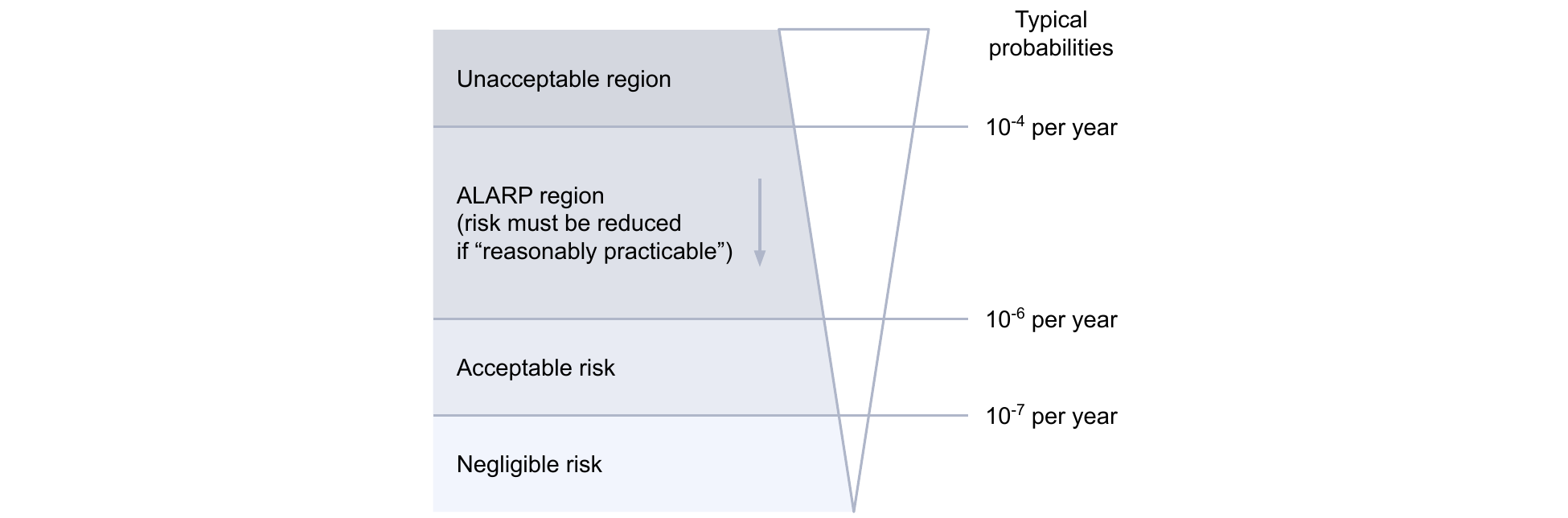}}
    \caption{The ALARP framework \protect\citep{Melchers2001-bf}}
    \label{fig:4}
\end{figure}

For example, many industries stack two risk thresholds: one threshold
above which risk is unacceptable and another one above which risk must
be ``as low as reasonably practicable'' (``ALARP''), also sometimes
referred to as ``as low as reasonably achievable'' (``ALARA'')
\citep{Linkov2011-cc,Melchers2001-bf}. The ALARP
framework, as illustrated in
\Cref{fig:4}, originated in the UK
Health and Safety at Work etc. Act 1974 and
has since been used in various industries worldwide, including the UK
and U.S. nuclear industry \citep{Hse1992-ft,Nrc2016-ey},
the U.S. aerospace industry \citep{Dezfuli2015-au}, and
the international maritime industry \citep{Imo2018-bo}.
Typically, risk is considered ALARP if the costs of further risk
reduction measures are ``grossly disproportionate'' to their benefits
\citep{Hse1992-ft}. The ALARP principle may ensure
continuous risk reduction efforts \citep{Aven2015-ug}.
However, it has also been criticized for being vague, leading to harmful
risk aversion, and stifling innovation
\citep{Melchers2001-bf,Oakley2020-sa}. Theoretically, any
other types of risk acceptance criteria could be applied at the risk
tier in the middle, allowing risk thresholds to be combined with the
other types of risk acceptance criteria mentioned in
\Cref{risk-thresholds}.

\subsection{Using risk thresholds to indirectly feed into
decisions}\label{using-risk-thresholds-to-indirectly-inform-decisions}

Risk thresholds can also be used to indirectly feed into decisions, such as
whether to deploy a model, by helping set capability thresholds.
U.S. nuclear regulators use risk thresholds to help determine adequate
safety measures
(\Cref{risk-thresholds}). In the
context of frontier AI, capability thresholds are emerging as a key
trigger for safety measures
(\Cref{capability-thresholds}).
Capability thresholds and corresponding safety measures could be
designed such that they would be estimated to keep risk below some risk
threshold. In this way, as capability thresholds directly feed into
high-stakes decisions, risk thresholds indirectly feed into decisions
(\Cref{fig:1}).

\begin{figure}[t!]
    \centering
    \makebox[\textwidth][c]{\includegraphics[width=1.1\linewidth]{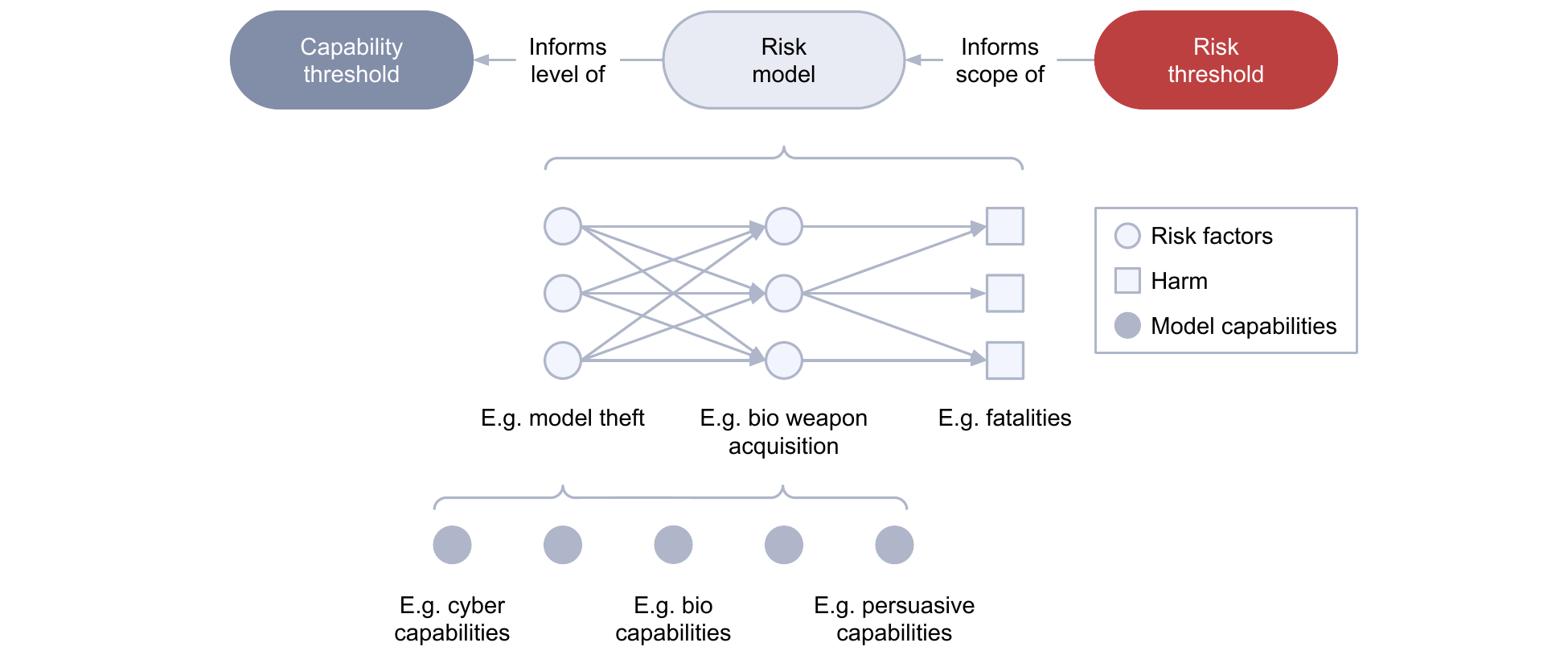}}
    \caption{Risk thresholds, for example via risk models, can help set capability thresholds}
    \label{fig:5}
\end{figure}

A helpful tool when using risk thresholds to help set capability
thresholds is ``risk models'' (see
\citealp{Google_DeepMind2024-qj}), also referred to as
``threat models'' \citep{Anthropic2023-bl}.\footnote{The
latter term is currently the most common, but may not be the most
suitable. It stems from a security context and may thus lead to a
narrow focus on security risks. Moreover, ``threat modeling''
encompasses more than outlining risk scenarios, in particular,
prioritizing among safety measures
\citep{Shostack2014-ij}. In standard risk management,
common terms to refer to risk models are ``fault trees'' and ``event
trees'' \citep{Barrett2017-ej}, ``attack trees'' as a
variation of fault trees for security risks
\citep{Salter1998-bh,Schneier2011-uj}, and ``causal
maps'' to depict non-linear relationships between risk factors
\citep{Iso2019-kp,Koessler2023-hi}.} Risk models
outline the pathways from risk factors to harm, or ``risk scenarios''
(see \citealp{OpenAI2023-tt}). They can be used to identify
model capabilities that may cause large-scale harm and safety measures
that may prevent such harm. For example, the capability to provide
instructions for the acquisition of biological weapons (dark circle) may
increase the risk of fatalities, economic damage, and societal
disruption (squares). Risk models may help identify the level of bio
capabilities at which the level of risk would exceed the risk threshold
unless safety measures are implemented
(\Cref{fig:5}).

In more detail, risk thresholds can be used to set capability thresholds
via risk models with the following steps. First, define risk thresholds.
We provide some guidance for doing so in
\Cref{how-to-define-ai-risk-thresholds}.
Second, develop risk models. Ideally, risk models are comprehensive,
meaning they contain all possible pathways from risk factors to harm.
However, developing comprehensive risk models is generally extremely
difficult \citep{Shostack2014-ij} and particularly so in
the case of a general-purpose technology like AI
(\Cref{arguments-against-using-risk-thresholds}).
Therefore, at least in the beginning, risk models may focus on a small number of key
risk scenarios. Third, identify model capabilities that would lead to
unacceptable risks as defined in the first step. This can draw on, for
example, the risk models developed in the second step, data gathered
about the occurrence of risk factors, near misses, and small-scale harm,
as well as methods like trend extrapolation and sensitivity analysis
\citep{Frey2002-ct}.

Frontier AI companies setting capability thresholds already aim to
identify the model capabilities that may lead to large-scale harm. In
particular, companies increasingly engage in risk modeling
\citep{Anthropic2023-bl,OpenAI2023-tt,Google_DeepMind2024-qj}.
Yet, when doing so, companies currently seem to mostly focus on the
\emph{possibility} that model capabilities may cause large-scale harm
rather than also considering the \emph{likelihood} of this happening.
Ignoring likelihood means ignoring a key component of risk and can lead
to overly restrictive capability thresholds, because other factors may
prevent harm from materializing, such as malicious actors not having
access to the model or society ramping up its defenses. At least one
company is planning on taking likelihood into account in the future
\citep{Anthropic2023-bl}.

\section{The case for AI risk
thresholds}\label{the-case-for-ai-risk-thresholds}

In this section, we argue that risk thresholds are a promising tool for
making high-stakes AI development and deployment decisions. Risk
thresholds may help align business conduct with societal concern; enable
consistent allocation of safety resources; ensure risk estimation
results are actually acted upon; prevent motivated reasoning regarding
what level of risk is acceptable; and avoid locking in premature safety
measures
(\Cref{arguments-for-using-risk-thresholds}).
We also discuss the most important objections to using AI risk
thresholds and how they might be overcome. In particular, estimating
risks from AI is extremely hard; AI is a dual-use, general-purpose
technology; risk thresholds may create an incentive to produce artificially
low risk estimates; and defining risk thresholds for AI involves handling
thorny normative trade-offs
(\Cref{arguments-against-using-risk-thresholds}).
Overall, we suggest that risk thresholds should be used indirectly inform high-stakes decisions by helping set, though not determine,
capability thresholds. Further, we suggest that risk thresholds may be used to
directly inform, though not determine, decisions. If and when our
ability to produce risk estimates improves, we can rely more on risk
thresholds
(\Cref{overall-suggestions-for-using-ai-risk-thresholds}).

\subsection{Arguments for using risk
thresholds}\label{arguments-for-using-risk-thresholds}

First and foremost, risk thresholds are focused on potential harms to
society and may thereby help align business conduct with societal
concern. Risk thresholds directly pertain to externalities: the
likelihood and severity of harm to individuals, groups, and society as a
whole. In contrast to compute thresholds and capability thresholds, risk
thresholds do not run into the issue of focusing on wrong proxies for
risk, such as harmless models or capabilities. As a result, risk
thresholds can help ensure companies only go ahead with risky
activities if the risk is acceptable to society.

Second, risk thresholds can enable consistent allocation
of safety resources. Risk thresholds can use the same units (e.g.
expected number of fatalities or amount of economic damage in USD) for
different risks (e.g. cyber and CBRN risks). As a result, risk
thresholds can be set at the same level for different risks. If done
well, this leads to a consistent allocation of safety resources. By
contrast, capability thresholds (set without the help of risk
thresholds) may inadvertently be set at different levels of risk for
different model capabilities (e.g. autonomy and persuasion
capabilities). This leads to an inconsistent allocation of safety
resources. However, we note that this benefit can also be achieved by
merely conducting risk estimates and consistently allocating safety
resources based on their results, without also setting risk thresholds.

Third, in contrast to merely conducting risk estimates, risk thresholds
can help ensure risk estimation results are actually acted upon.
When using risk thresholds to directly feed into decisions, risk thresholds
link the results of risk estimates to decisions (in the simplest
version, go or no-go). When using risk thresholds to indirectly feed into
decisions, risk estimation results are ``enshrined'' in capability
thresholds, which in turn are integrated in decision rules (again, in
the simplest version, go or no-go). In both ways, risk thresholds may
help avoid situations where risk estimates are produced but not acted
upon.

Fourth, risk thresholds may prevent companies from engaging in motivated
reasoning regarding what level of risk is acceptable. Companies have
strong incentives to argue that risk estimates are acceptable \emph{in
hindsight}. Risk thresholds can prevent this by determining criteria for
what level of risk is acceptable \emph{in advance}.\footnote{Note that
for risks where AI exacerbates a baseline risk, such as the current
risk of cyberattacks on critical infrastructure, this baseline risk
needs to be estimated before risk thresholds can be defined. However,
even in these cases, risk thresholds are defined before the increase
in risk caused by the AI development or deployment decision is
estimated. This means that risk thresholds for any risk are defined
before the increase in risk caused by AI is estimated.} Yet, on the
flipside, risk thresholds increase the incentive for companies to
provide lower risk estimates in the first place. We discuss this concern
below
(\Cref{arguments-against-using-risk-thresholds}).

Fifth, risk thresholds are future-proof and may help avoid locking in
premature safety measures. Given that AI is still evolving (and rapidly
so), regulators face the question of how prescriptive their requirements
should be. Here, risk thresholds can provide a way out, as they do not
require regulators to specify which safety measures companies need to
implement. Instead, if regulators mandate risk thresholds to directly or
indirectly feed into decisions (leaving it to companies to set capability
thresholds), regulators put the burden on companies to find ways to
reduce risk and can even continuously incentivize companies to innovate
on safety measures, which may lower costs and result in more effective
safety measures
\citep{Decker2018-kd,Schuett_et_al_undated-wx}. On the
other hand, if regulators mandate risk thresholds, they need a lot of
effort and expertise to verify compliance
\citep{Decker2018-kd,Schuett_et_al_undated-wx}.
Regulators need to check whether the risk estimates that companies have
produced are sound, which involves a case-by-case analysis of companies'
risk estimates. But regulators can require companies to provide them
with detailed information about their risk estimates and the reasoning
behind them, for example, through established tools like safety cases
\citep{Bishop2000-dj,Buhl_et_al_undated-up,Kelly1998-ga}.
Still, given the incentive for companies to provide low risk estimates
(see previous paragraph), verifying compliance with risk thresholds may
necessitate regulators to conduct their own risk estimates.

\subsection{Arguments against using risk
thresholds}\label{arguments-against-using-risk-thresholds}

The key argument against using risk thresholds is that risk estimation
is extremely hard for risks from frontier AI development and deployment.
Using risk thresholds requires estimating the level of risk. In general,
estimating risks from complex technological systems is hard
\citep{Apostolakis2004-fm}. This issue is aggravated in
the case of frontier AI. There is little data from past incidents,
meaning risk estimates mostly have to draw from modeling and expert
judgment, which are less reliable. In general, risk estimation struggles
with low-probability, high-impact events and ``unknown unknowns'', which
may be features of many risks from frontier AI. On top of that,
understanding of how AI systems work and why they fail is poor, risk
taxonomies and risk models are underdeveloped, and relevant information
is split between companies and regulators -- companies have knowledge of
AI capabilities and usage, while regulators possess intelligence data,
including about societal vulnerabilities and the number, capacities, and
incentives of malicious actors. It might be possible to alleviate these
issues, for instance, by improving risk estimation methodologies and
gathering data about the occurrence of risk factors, near misses, and
small-scale harm \citep{Schuett_et_al_undated-wy}. Nevertheless, the lack
of reliable risk estimates currently is the main limitation of risk
thresholds. The more strongly high-stakes decisions rely
on risk thresholds, the more reliable these risk estimates should be.

Second, and relatedly, a common objection to using risk thresholds in 
frontier AI
regulation is that foundation models, similar to electricity, are a
dual-use or general-purpose technology that can be used in a tremendous
number of ways and have a tremendous number of consequences that are
both impossible to foresee and not the responsibility of frontier AI
companies to prevent. This is a valid concern. However, this is a common
issue in tort and criminal law, where mere causation is not enough \citep{Wright1985-qu}. Likewise, in this
context, this issue does not refute risk thresholds in general but means
that regulators need to specify which effects are in scope (see also
\Cref{type-of-risk}). Where to draw
this line is a strategic decision that involves a variety of
considerations (including economic, geopolitical, fairness, and safety
considerations). Relevant qualitative criteria may be what type and
amount of harm is at stake and whether intervention at later stages can
be expected to be sufficiently effective
\citep{Anderljung2023-xo}. Based on these criteria,
imposing risk thresholds on frontier AI companies may be especially
warranted for scenarios where single events cause large-scale harm and
where no downstream developers are involved who could be held
accountable instead.

Third, risk thresholds may create an incentive to produce artificially low 
risk estimates. If risk thresholds are used to directly
feed into decisions, they establish a clear link between risk
estimates and high-stakes decisions, making the implications of risk
estimates immediately obvious. If risk thresholds are used to indirectly feed into decisions, the conditional risk estimates for different model capabilities have a less clear, but still perceivable impact on capability thresholds and thus decisions. Companies can take advantage of the
uncertainty and subjectivity of the risk estimation process to produce the results they desire, with an added veneer of
plausibility. To address this concern, regulators can verify companies'
risk estimates or mandate procedural requirements, such as that
companies must involve more, diverse, and external assessors or that
they break down risks into multiple events and ask assessors to estimate
the risk of the individual events only. For example, instead of asking
each assessor to estimate the increase in risk from biological attacks,
companies could ask separate assessors to estimate the increases in risk
regarding ideation, acquisition, magnification, formulation, and release
of biological weapons \citep{Patwardhan2024-ep}.

Fourth, it can be very difficult to decide what level of risk is
acceptable \citep{Aven2015-ug}. In particular, this
decision involves making thorny normative judgments such as how much to
value a human life \citep{Reid2000-re,Vanem2012-vs}, how
much to value future generations \citep{Aven2012-ue} or
the environment \citep{Vanem2012-vs}, and how cautious to
be in the face of high uncertainty \citep{Klinke2002-bm}.
While making these decisions can be challenging for a single person, it
will be even harder for different people or society as a whole to agree
on the choice. Yet these decisions are currently being made implicitly
through company development and deployment decisions. The fact
that defining risk thresholds will be tough provides an argument for
getting started sooner rather than later, such that
important discussions and investigations can take place with sufficient
time and rigor. We aim to help start this process by providing some guidance on
how to define risk thresholds in
\Cref{how-to-define-ai-risk-thresholds}.

\subsection{Overall suggestions for using AI risk
thresholds}\label{overall-suggestions-for-using-ai-risk-thresholds}

Risk thresholds should be used to indirectly feed into high-stakes decisions, and may
additionally be used to directly feed into such decisions. The respective
benefits and limitations of risk thresholds and capability thresholds
mean that risk thresholds should complement, rather than replace, capability thresholds. 
Risk thresholds are directly focused on potential harms to society. However,
they rely on risk estimates, which still have methodological limitations
and involve substantial uncertainties, whereas capability thresholds
rely on model evaluations, whose results are significantly less
uncertain. Therefore, the key use case for risk thresholds should be to
help set capability thresholds, ensuring that capability thresholds and
corresponding safety measures, if followed, keep risk to an acceptable
level. Additionally, using risk thresholds to directly feed into
high-stakes decisions is helpful if capability thresholds
miss the mark or become outdated. In conclusion, the two use cases of
risk thresholds are not mutually exclusive but can make up for the
limitations of the other. Hence, they should be applied in combination
(\Cref{fig:1}).

Nevertheless, as long as risk estimates are not reliable, risk
thresholds in both of their use cases should not determine, but only
inform, high-stakes decisions. The difficulty of producing reliable risk
estimates is the strongest reason against using risk thresholds as the sole basis for a strict decision-rule for whether to go ahead or for where to set capability thresholds. It
means that, in both cases, risk thresholds should currently only be used as one among a number of considerations. We provide some concrete examples for how risk thresholds can directly and indirectly inform, rather than determine, high-stakes decisions below. At the same time, to facilitate greater reliance on risk
thresholds in the future, regulators and companies should invest in
improving risk estimation methodology, gain experience in conducting
risk estimates, investigate how much they can rely on them, and
gather data about the occurrence of risk factors, near misses, and
small-scale harm.

Concretely, when using risk thresholds to directly inform high-stakes
decisions, they should be used among a number of other considerations, such as
capability thresholds (which may or may not have been set with the help
of risk thresholds). Moreover, many considerations unrelated to societal risk will
come into play, including the company's appetite for business risks (e.g. liability risk or reputational risk) and various strategic
considerations (e.g. whether a competitor is likely to release
a similar model soon) (see \citealp{Iso2019-kp}). Beyond that, could inform, rather than determine, decisions in that, if they are crossed, the board
of directors or the competent regulator would have to be notified. The notified
actor could potentially also be allowed to veto the decision or be
required to give permission to proceed. Another option is that exceeding
a risk threshold would trigger a requirement to conduct an extra suite
of in-depth model evaluations, which may have to include third-party
evaluators.

When using risk thresholds to indirectly inform decisions by helping set
capability thresholds, other important considerations include expert judgment and 
safe design principles. Safe design involves, for instance, principles like
redundancy, defense in depth, loose coupling of components to avoid
cascading failures, separation of powers between decision-makers, and
fail-safe design ensuring that systems fail gracefully
\citep{Dobbe2022-bz,Leveson2016-sa,Perrow1999-nj,Reason1990-lf}. 
The company could also find that risk estimates suggest that one of its capability thresholds does not keep risk to an acceptable level, but nonetheless not change the capability threshold because the relevant model capabilities present substantial benefits or because they have good reasons to believe the risk estimates are unreliable. Moreover, the capability thresholds, set with the help of risk thresholds, may only inform, rather than determine, high-stakes decisions.

\section{How to define AI risk
thresholds}\label{how-to-define-ai-risk-thresholds}

In this section, we propose a framework that consists of important
considerations for defining AI risk thresholds. We did not find good
general guidance for how to define risk thresholds. Thus, we conducted a
non-systematic review of risk thresholds in various industries and
jurisdictions, including aviation, nuclear, aerospace, maritime, and
transportation and storage of hazardous materials, and tried to identify
common ground on the most important considerations. Before regulators or
companies can answer the question of what level of risk is acceptable,
they need to decide which type of risk the threshold refers to
(\Cref{type-of-risk}). Next, when
determining the acceptable level of risk, they need to handle three
difficult normative trade-offs: how to weigh potential harms and
benefits, to what extent should mitigation costs be taken into account, and 
how to deal with uncertainty regarding all of the aforementioned
(\Cref{level-of-risk}).

\subsection{Type of risk}\label{type-of-risk}

\begin{figure}
    \centering
    \makebox[\textwidth][c]{\includegraphics[width=1.15\linewidth]{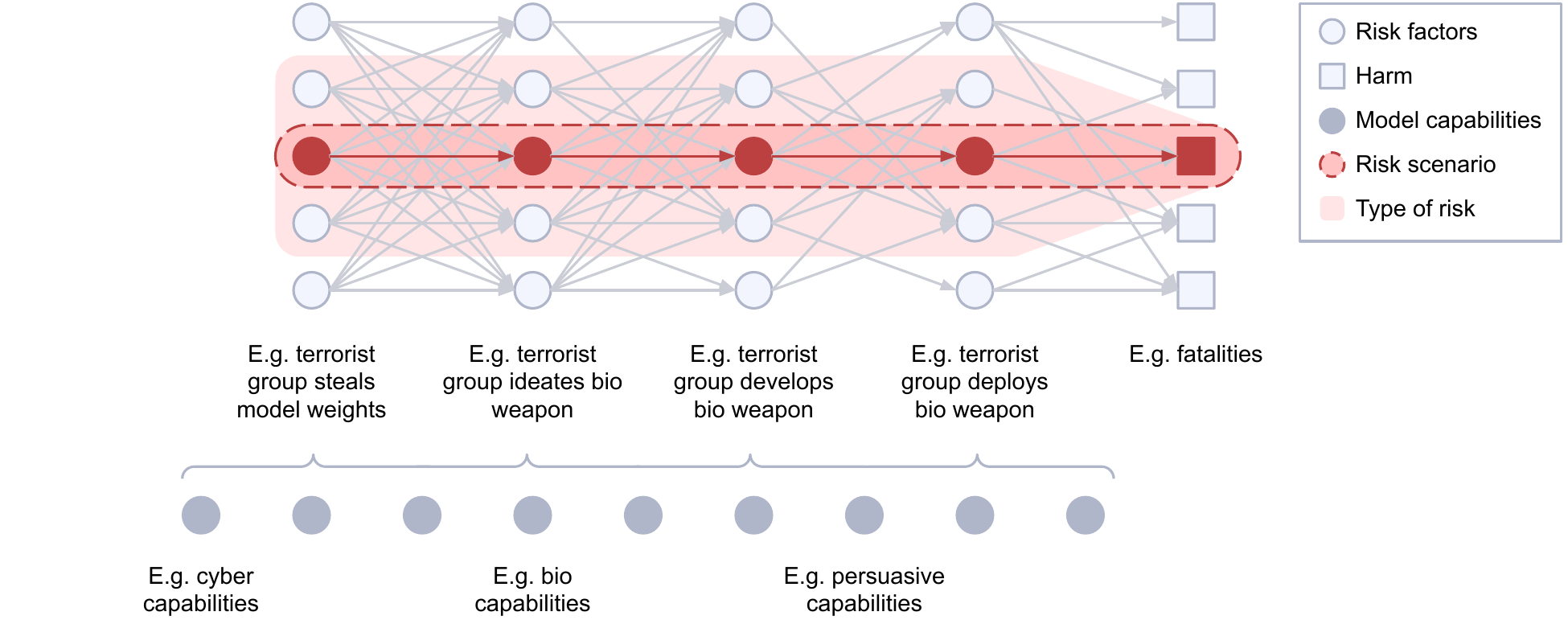}}
    \caption{Representation of a linear risk model consisting of many risk scenarios}
    \label{fig:6}
\end{figure}

Every risk threshold is set for a specific ``type of risk''. This term
does not have a standard definition. In general, a type of risk seems to
mean a group of risk scenarios that have similar impact, origin, or
other characteristics, and may also be referred to as ``area of risk''
or ``category of risk''. For example, common types of business risks
include financial, legal, reputational, operational, and strategic
risks. When it comes to risks from AI, types of risks could be
distinguished by type of harm (e.g. fatalities, injuries, and economic
damage)\footnote{Regulators and companies may want to begin with setting
risk thresholds for types of harm that are relatively easy to measure
(e.g. fatalities). However, types of harm that are harder to measure,
such as discrimination, disinformation, or societal disruption, should
not be neglected for this reason. But they require additional effort,
because regulators and companies need to develop suitable metrics
first. For a first effort in this regard, see
\citet{Solaiman2023-yt}.} and potentially additionally
by the domain or modality of occurrence of that harm (e.g. for
fatalities, this could mean distinguishing between fatalities that stem
from biological attacks, chemical attacks, and cyberattacks on critical
infrastructure)
(\Cref{fig:6}).\footnote{Additionally
distinguishing by domain or modality of occurrence of harm is not
necessary but may allow setting more quantitative risk thresholds for
types of risks where more data or better risk estimation methodologies
exist. It may also make it easier to use risk thresholds to help with
setting capability thresholds, which often focus on capabilities
relevant for a particular domain (e.g. bio capabilities). For
regulators, it can also make risk estimation easier, because
information about different types of risks may be located within
different government departments.}

The choice of how many risk scenarios are in scope may affect where to
set the risk threshold. All else equal, the fewer risk scenarios
included, the lower, i.e. more strict, the risk thresholds should be
because more fine-grained types of risks constitute a smaller fraction
of the overall risk. For example, the U.S. aerospace industry used to
have separate risk thresholds of $30\times 10^{-6}$ for the
level of risk from each of the three main risk scenarios during rocket
launch (explosive debris, toxic release, and blast overpressure). To
simplify the licensing process, the industry switched to a risk
threshold of $1\times 10^{-4}$ for the level of risk from all
three risk scenarios combined \citep{Faa2016-tb}. Note
that the overall threshold remained about the same ($3 \times 30 \times
10^{-6} \approx 1 \times 10^{-4}$).

Many regulators also use separate risk thresholds for different numbers
of fatalities. Two very common risk thresholds across many industries
and jurisdictions are ``individual risk thresholds'' and ``societal risk
thresholds''. While definitions vary, individual risk thresholds usually
refer to the risk of death of individuals, while societal risk
thresholds refer to the risk of death of groups of people from a single
event (e.g. \citealp{Hse2001-lo,Iaea2005-se,Imo2018-bo}).
In short, regulators often use separate risk thresholds for risk
scenarios where a single person dies and those where several people die.

The previous discussion can be considered to concern a risk threshold's
material scope -- in addition, the temporal scope and territorial scope
for harm to occur need to be defined. All else equal, the shorter the
time period taken into account for harm to materialize, the lower, i.e.
more strict, the risk thresholds should be, because shorter time periods
represent a smaller fraction of the overall risk. While longer time
periods are more comprehensive, shorter time periods are easier to
assess. For example, aviation has risk thresholds per flight-hour
\citep{Icao2018-hl}, whereas the nuclear industry defines
risk thresholds per reactor-year \citep{Iaea2005-se}. In
the case of AI, a temporal scope of 12 months may align well with the
yearly business cycle. However, developing biological weapons, for
instance, may take several years and would not be in scope in this case.
Similar considerations apply with regard to where the harm occurs. For
example, in the U.S. nuclear industry, the individual risk threshold
considers individuals within 1 mile of the power plant, whereas the
societal risk threshold considers the population within 50 miles of the
power plant \citep{Nrc1983-oo}. In the case of AI, the
territorial scope may need to be unrestricted, because frontier AI
companies provide their services globally, and harm may thus occur
anywhere in the world. For instance, cyberattacks can target any system,
especially if it is connected to the internet.

There also need to be rules for what type of causation is in scope; for
example, second-order effects may be excluded. At the very least, the
model needs to be causal for the harm. Causation can be established via
the ``but-for test'' from law \citep{Hart1985-yr}: ``but
for the model, would the harm have occurred?'' But mere causation may
not be sufficient for practical reasons, because it would include a
tremendous number of cases where the activity contributes marginally to
the occurrence of harm (see also
\Cref{arguments-against-using-risk-thresholds}).
Therefore, for example, a risk threshold could focus on first-order
effects, that is harms directly stemming from the AI development,
deployment, or use. Based on this example definition, harm to users or
harm caused by malicious actors would be in scope, whereas harm to
workers that are displaced by AI systems would not be in scope. However,
we highlight that the lines can be blurry, and clear rules need to be
established.

Finally, for risks where AI exacerbates a baseline risk (e.g.
cyberattacks) as opposed to creating a new risk (e.g. rogue AI
scenarios), it will usually be preferable for risk thresholds to refer
to the increase in risk caused by AI, i.e. ``marginal risk'', rather
than the total level of risk \citep{Kapoor2024-ss}. Note
that the increase in risk should still be expressed in absolute, not
relative, terms: a 5\% increase in deaths from heart attacks is far
worse than a 5\% increase in deaths from shark attacks. However, for
many risks from AI it is unclear what should be the relevant baseline
risk -- the level of risk with or without current AI systems, and
whether the former includes AI systems by the company itself or only AI
systems by its competitors. It is also unclear whether and, if so, how
risk estimates for risk thresholds should take into account the increase
in risk caused by expected AI systems from competitors. If they did,
that could create a situation where each frontier AI company behaves
recklessly in part because it reasons that its competitors will behave
recklessly. What constitutes the relevant baseline risk needs to be
clearly defined.

\subsection{Level of risk}\label{level-of-risk}

There seem to be three main ways used to determine the acceptable level
of risk: building on peoples' revealed preferences, copying what other
industries do, and doing cost-benefit analysis
\citep{Philipson1983-la,Reid2000-re}. Some regulators
have reviewed the level of risk that people accept through engaging in
common activities like driving (e.g. \citealp{Hse1992-ft}).
Other regulators have reviewed the level of risk that society accepts in
other industries with comparable benefits (e.g.
\citealp{Nrc1983-oo} -- comparing nuclear to coal, ``the
competing form of generating electricity'').

Most regulators appear to have reviewed and copied the risk thresholds
already used in other industries or jurisdictions. As a potential
result, many regulators use the same individual risk threshold of $1\times 10^{-6}$ per fatality and year (e.g.
\citealp{Hse2001-lo,Iaea2005-se,Imo2018-bo}). However,
societal risk thresholds seem to vary more strongly. Among industries
and jurisdictions, the acceptable risk threshold for 1,000 fatalities
ranges from a likelihood of $1\times 10^{-6}$ to
$1\times 10^{-11}$ per year
\citep{Ehrhart2020-dk}. This study finds, for example,
that the UK Health and Safety Commission sets risk thresholds for the
transport of dangerous substances, deeming $1\times 10^{-4}$
unacceptable and $1\times 10^{-6}$ acceptable. By contrast, the
survey finds that the Swiss Federal Office for the Environment sets risk
thresholds for fixed installations and tunnels, deeming
$1\times 10^{-9}$ unacceptable and $1\times 10^{-11}$
acceptable.

Few regulators appear to have conducted systematic cost-benefit analysis
to determine the acceptable level of risk. A notable exception seems to
be the maritime industry
\citep{Emsa2015-ai,Imo2018-bo}. Choosing the
acceptable level of risk in a systematic way is extremely difficult
\citep{Hse1992-ft}. However, given that AI may not be
comparable to any other industry in terms of the benefits it might
generate, this may be the necessary approach. In the following, we
provide some initial guidance on the three key normative trade-offs that
need to be handled in a systematic cost-benefit analysis: how to weigh
potential harms and benefits, to what extent to take into account
mitigation costs, and how to deal with large amounts of uncertainty.

The key question when determining the acceptable level of risk for an
activity is how to weigh the many potential harms against the benefits
that may come from it \citep{Hubbard2020-la}.\footnote{One
often reads that benefits should not be taken into account above the
unacceptable risk threshold (e.g. \citealp{Hse2001-lo}).
But that guidance seems to refer to the moment when the threshold is
\emph{used}. When the threshold is \emph{defined}, benefits should
always be taken into account.} Greater benefits can be accounted for
by setting higher, i.e. less strict, thresholds.\footnote{A key decision
that needs to be made is which benefits are in scope -- this raises
parallel questions to which harms are in scope, so we refer to that
discussion (\Cref{type-of-risk}).}
But can the benefits of scientific advances be weighed against the harms
of discrimination or disinformation? This is extremely challenging (see
\Cref{arguments-against-using-risk-thresholds}).
For example, regulators in the maritime industry have developed a target
societal risk/benefit ratio, the amount of societal benefit
necessary to outweigh the risk of a single fatality. They derived the
target societal risk/benefit ratio from aviation -- because aviation has
``good statistical data'' and an ``excellent safety record'' --
estimating the benefits via company revenues. They then apply this
target societal risk/benefit ratio to the maritime industry
\citep{Emsa2015-ai,Imo2018-bo}. A key issue
with using this approach for AI is that many societal benefits, such as
fundamental scientific advances, may not be reflected in company
revenue.

The second key trade-off when determining the acceptable level of risk
is to what extent to take into account the costs of reducing risks, in
terms of money, time, or effort. Greater mitigation costs can be
accounted for by setting higher, i.e. less strict, thresholds.
Alternatively, as discussed in
\Cref{using-risk-thresholds-to-directly-inform-decisions},
a common approach in other industries is to set two risk thresholds: one
above which risk is unacceptable regardless of mitigation costs, and one
above which risk must be ``as low as reasonably practicable'' or
``ALARP''; that is, risk must be reduced until the costs would be
``grossly disproportionate'' to the benefits of further risk reduction
\citep{Hse1992-ft}. This means mitigation costs do not
influence the acceptable level of risk, but above some level of risk
they influence what must be done if the threshold is crossed.

The third key trade-off when determining the acceptable level of risk is
how to set the expected ratio of false negatives to false positives.
Estimates of harms, benefits, and mitigation costs will involve large
amounts of uncertainty. An approach that is more risk tolerant and
therefore more concerned about benefits and mitigation costs, i.e. false
positives (either due to the risk threshold accidentally being set too
low or the level of risk wrongly being estimated to be above the
threshold), leads to higher, i.e. less strict, risk thresholds. The more
the risk threshold should be risk averse and reflect concern about
harms, i.e. false negatives (either due to the risk threshold
accidentally being set too high or the level of risk wrongly being
considered below the threshold), the lower, i.e. more strict, the risk
thresholds should be to generate a ``margin of safety''. It seems
prudent to have a margin of safety that is larger the more consequential
and irreversible the type of harm at stake (e.g. this applies more to
fatalities than to injuries). Some regulators also choose to be more
risk averse the larger the harm at stake. For example, the Dutch nuclear
industry sets its societal risk threshold at a probability of $1\times 10^{-5}/\mbox{N}^2$ for $10 \times \mbox{N}$ fatalities
per year \citep{Anvs2020-rn}. The division by
$\mbox{N}^2$ instead of N means a steeper slope of the risk
thresholds and reflects aversion to large accidents (the acceptable
probability decreases exponentially instead of linearly with the number
of fatalities increasing).

Generally, regulators have more legitimacy and better incentives to
define socially desirable thresholds than companies do, in particular if
companies' activities may cause externalities to society
\citep{Abrahamsen2012-pn}. The public safety and security
risks that may stem from frontier AI systems are such externalities.
Therefore, ideally regulators, but at least companies, should define
risk thresholds.

\section{Conclusion}\label{conclusion}

This paper has made four main contributions. First, we have clarified
the concepts of risk thresholds, capability thresholds, and compute
thresholds, arguing that they not only rely on different metrics, but
should also serve different functions. Second, we have made the case
that risk thresholds are a promising tool for frontier AI regulation to the extent that the 
reliability of risk estimates can be improved. Third, we have argued that risk
thresholds should be used to indirectly inform high-stakes decisions by
helping set, but not determine, capability thresholds, and may also be used to directly
inform, but not determine, high-stakes decisions. Fourth, we have developed
initial guidance for defining AI risk thresholds.

Many questions around risk thresholds for high-stakes AI development and
deployment decisions warrant much further research. We highlight some
questions that seem especially important. Fundamentally, advancing the
risk estimation methodology is of utmost importance if regulators and
companies want to rely more on risk thresholds for high-stakes
decisions. In that regard, developing risk taxonomies and risk models,
gaining experience with risk estimation methods, as well as gathering
data about the occurrence of risk factors, near misses, and small-scale
harm may be among the most useful ways forward (see \citealp{Schuett_et_al_undated-wy}). The details of how to use risk thresholds to determine
adequate capability thresholds and corresponding safety measures also
need to be explored further. In this regard, companies and regulators
may be able to learn from the U.S. nuclear industry (see
\citealp{Nrc2021-fe}). Last but not least, the acceptable
levels of risk for different types of risks need to be defined. To do
so, regulators could conduct comparative studies of thresholds in other
industries or systematic cost-benefit analyses.

Managing the risks from frontier AI systems is an important and urgent
challenge. Frontier AI companies need to improve their risk management
practices and should use risk thresholds to help set capability
thresholds. Over time, high-stakes development and deployment 
decisions should be directly informed by risk thresholds set by regulators. 
To this end, we need a discussion about what level of risk we, as a society, are 
willing to accept.

\section*{Abbreviations}\label{abbreviations}

\renewcommand{\arraystretch}{1.2}
\begin{tabular}{@{}ll}
    AI & Artificial intelligence \\
    ALARA & As low as reasonably achievable \\
    ALARP & As low as reasonably practicable \\
    ANVS & Dutch Authority for Nuclear Safety and Radiation Protection \\
    BEIS & UK Department for Business, Energy and Industrial Strategy \\
    CCPS & Center for Chemical Process Safety \\
    COSO & Committee of Sponsoring Organizations of the Treadway Commission \\
    DSIT & UK Department for Science, Innovation and Technology \\
    EMSA & European Maritime Safety Agency \\
    ESA & European Space Agency \\
    EUROCONTROL & European Organisation for the Safety of Air Navigation \\
    FAA & U.S. Federal Aviation Administration \\
    HSE & UK Health and Safety Executive \\
    IAEA & International Atomic Energy Agency \\
    ICAO & International Civil Aviation Organization \\
    IEC & International Electrotechnical Commission \\
    IMO & International Maritime Organization \\
    ISO & International Organization for Standardization \\
    NASA & U.S. National Aeronautics and Space Administration \\
    NIST & U.S. National Institute of Standards and Technology \\
    NFPA & U.S. National Fire Protection Association \\
    NRC & U.S. Nuclear Regulatory Commission \\
    ONR & UK Office for Nuclear Regulation
\end{tabular}

\section*{Acknowledgments}\label{acknowledgments}

We are grateful for valuable input from the following individuals,
listed in alphabetical order by surname: Jide Alaga, Bill
Anderson-Samways, Anthony Barrett, Caroline Baumoehl, Samuel Bowman, Ben
Bucknall, Marie Buhl, Francisco Carvalho, Alan Chan, Noemi Dreksler, Ben
Garfinkel, James Ginns, John Halstead, Jonathan Happel, Lennart Heim,
Samuel Hilton, Holden Karnofsky, Patrick Levermore, Eli Lifland, David
Lindner, Sebastien Krier, Yannick Muehlhaeuser, Malcolm Murray, Aidan
O'Gara, Cullen O'Keefe, Alex Rand, Luca Righetti, Josh Rosenberg, Gaurav
Sett, Rohin Shah, Merlin Stein, Christopher Phenicie, Hjalmar Wijk, Zoe
Williams, and Peter Wills. All views and remaining errors are our own.

\section*{Declarations}\label{declarations}

One author's spouse holds equity in a frontier AI
company. Apart from that, the authors have no relevant financial or
non-financial interests to disclose.

\bibliographystyle{apacite}
\bibliography{ms}

\end{document}